\documentclass{aastex631}

\usepackage{multirow}
\usepackage{amsmath}

\newcommand{\htwoco}{H\ensuremath{_2}CO}
\newcommand{\rrl}{H\ensuremath{_{110\mathrm{\alpha}}}}
\newcommand{\bh}{BGPS-\htwoco{}}
\makeatletter
\DeclareRobustCommand{\hi}{%
  \mbox{H\check@mathfonts\fontsize\sf@size\z@\selectfont I}%
}
\DeclareRobustCommand{\hii}{%
  \mbox{H\check@mathfonts\fontsize\sf@size\z@\selectfont II}%
}
\makeatother

\newcommand{\highlight}[1]{\textbf{#1}}

\begin{document}

\title{
Rethinking Habitability using Biogenic Precursors: Formaldehyde in Millimeter Molecular Clouds of the Inner Galaxy}

\author[0009-0009-4694-5097]{N. B. Baharin}
\affiliation{Department of Physics, Faculty of Science\\
Universiti Malaya\\
50603 Kuala Lumpur, Malaysia}

\author[0009-0008-8190-1607]{A. A. Nazri}
\affiliation{Department of Physics, Faculty of Science\\
Universiti Malaya\\
50603 Kuala Lumpur, Malaysia}

\author[0000-0002-0107-8980]{Z. Rosli}
\affiliation{Centre of Foundation, Language and Malaysian Studies\\
Universiti Malaya-Wales\\
50480 Kuala Lumpur, Malaysia}
\affiliation{Department of Physics, Faculty of Science\\
Universiti Malaya\\
50603 Kuala Lumpur, Malaysia}

\author[0000-0002-7149-0997]{Z. Z. Abidin}
\affiliation{Department of Physics, Faculty of Science\\
Universiti Malaya\\
50603 Kuala Lumpur, Malaysia}

\author[0000-0001-6513-2432]{H. A. Tajuddin}
\affiliation{Department of Chemistry, Faculty of Science\\
Universiti Malaya\\
50603 Kuala Lumpur, Malaysia}

\author{J. Esimbek}
\affiliation{Xinjiang Astronomical Observatory\\
Chinese Academy of Sciences\\
Urumqi 830011, People's Republic of China}

\author{D. Li}
\affiliation{Xinjiang Astronomical Observatory\\
Chinese Academy of Sciences\\
Urumqi 830011, People's Republic of China}

\author{X. Tang}
\affiliation{Xinjiang Astronomical Observatory\\
Chinese Academy of Sciences\\
Urumqi 830011, People's Republic of China}



\begin{abstract}
We present a comprehensive study of formaldehyde (\htwoco{}) absorption and radio recombination line (\rrl{}) emission in 215 molecular clouds from the Bolocam Galactic Plane Survey (BGPS), observed using the Nanshan 25-m radio telescope. \htwoco{} was detected in 88 sources (40.93\%) with 59 being new detections, while \rrl{} emission was found in only 11 sources (5.12\%), all coincident with \htwoco{} absorption. There exists a correlation of \htwoco{} fluxes with millimeter fluxes below a $3~\mathrm{Jy}$ threshold and an increased dispersion above it, suggesting the sub-CMB cooling of \htwoco{}. Cross-matching with kinematic distance catalogs revealed \htwoco{} spanning galactocentric distances from $0.216$ to $10.769~\mathrm{kpc}$, with column densities ranging from $7.82\times10^{11}$ to $6.69\times10^{14}~\mathrm{cm}^{-2}$. A significant inverse correlation was observed between \htwoco{} detection fraction and galactocentric distance, suggesting enhanced star-forming activity closer to the Galactic Center. These findings challenge traditional Galactic Habitable Zone (GHZ) models by demonstrating the presence of biogenic precursors in the inner Galaxy, shielded within dense molecular clouds. Our results underscore the importance of incorporating chemical tracers like \htwoco{}, alongside physical constraints, to refine the boundaries of the GHZ and advance the research of prebiotic chemistry in the Milky Way.
\end{abstract}



\section{Introduction} \label{sec:intro}

\cite{Tucker1981} initially presented how supernovae and gamma ray bursts (GRBs) could sterilize regions of the galaxy by exposing planets to intense radiation, and so laid the groundwork for the concept of an annular Goldilocks area in the disk of a galaxy that is ``just right'' for life to form. \cite{Gonzalez2001, Gonzalez2005} then further refined the concept of a Galactic Habitable Zone (GHZ) by incorporating factors like metallicity, stellar density and the frequency of catastrophic events. Now, GHZ is being increasingly utilized for the discovery of exoplanets \citep{Nari2025}, explorations of habitability of M-dwarfs in the galaxy \citep{Sagear2023} and likelihood of terrestrial planets to form around a parent star \citep{Gowanlock2011}, thus making essential the efforts in constraining the inner and outer radii of the GHZ. The accepted borders of the GHZ of Milky Way are within approximately $7$ to $10~\mathrm{kpc}$ from the Galactic Center, with 75\% of the stars older than the Sun, aged $8$ to $4~\mathrm{Gyrs}$ \citep{Lineweaver2004} with its inner boundary being constrained by radiation hazards and the outer line being limited by insufficient metallicity for planet formation \citep{Gonzalez2001}. \cite{Spitoni2014} widened the GHZ acreage to $8$ to $12~\mathrm{kpc}$ from the Galactic center, peaking at 10 kpc when taking into account the higher number of stars with habitable environments that formed in the inner regions due to weaker radial gas inflows but it is followed by the postulation that the GHZ started as a narrow band at large radii, and then settles into a range of $2$ to $13~\mathrm{kpc}$ as exhibited by a three-dimensional, asymmetric modeling of the temporal evolution of GHZ \citep{Forgan2016}. In limiting the outer boundary of the GHZ, N-body simulation models of the Galactic Habitable Zone by \cite{Vukotic2016} predicted stellar particles with mass of $5 \times 10^4~M_{\odot}$ inhabit regions of $\sim 16~\mathrm{kpc}$ for most of their lifetimes and so implies areas in the $10$ to $15~\mathrm{kpc}$ range to be most feasible in harboring habitable systems. Conversely, Monte Carlo modeling methods by \cite{Gowanlock2011} interestingly pull back the inner radius of GHZ to a minimum of $R < 4.1~\mathrm{kpc}$ and above the disk's midplane at $z \approx 1.5~\mathrm{kpc}$, as this is the region with the greatest fraction of stars with habitable planets when considering sufficient planet-forming levels of metallicity and stellar density that negate the impact of supernova sterilizations.  

However, it was also proposed that the inner radius of GHZ could even encompass the Galactic bulge \citep{Balbi2020} as it considers the broad metallicity distribution in the inner $2~\mathrm{kpc}$ of the bulge, ranging from $Z = -0.1$ to $0.4$ and the time between projected SN-induced mass extinction events, around $40$ and $110~\mathrm{Gyr}^{-1}$. \cite{Prantzos2007} modeled stellar distribution containing solar-like systems in five epochs throughout the Galaxy's evolution and found that all metallicity-dependent probabilities peak early in the inner disk and due to the inside-out formation pattern of the Milky Way, spread towards the outer limits of the galaxy and in the time it took to do so, the absolute probability value of surviving SN explosions would be substantial in the inner disk, and so cursorily opined the entire Milky Way disk could be a habitable zone.

Another method of constraining the boundaries of a GHZ is by chemically mapping Milky Way regions for biogenic molecules to identify areas in which the necessary reactions ostensibly occur in order to produce complex organic molecules (COMs) that can then sequentially give rise to amino acids and proteins, the building blocks for life. Contrary to its name, COMs are only defined as complex in context of the environment in which these species interact such as the interstellar medium (ISM) and star forming regions (SFRs) but are fundamentally compounds containing at least six atoms with a backbone of carbon, and are usually found in dense, cold molecular clouds that act as reservoirs and natural shields, allowing their formation and preservation \citep{Herbst2022}. As noted by \cite{Ziurys2018}, interstellar molecules e.g. formaldehyde (\htwoco{}) and methanol (CH$_3$OH) are critical precursors to life, and their distribution throughout the galaxy by way of a molecular lifecycle mechanism should be a key factor in defining the GHZ. Prebiotically plausible molecules have been detected as far as $23.6~\mathrm{kpc}$ (Edge Cloud 2), low metallicity and enhanced cosmic ray ionization rates notwithstanding \citep{Ruffle2007}. At distances $> 16~\mathrm{kpc}$, the far outer Galaxy is characterized by lower gas and stellar densities, weaker interstellar radiation fields, and fewer supernova remnants to trigger star formation, and yet a significant number of molecular clouds are observed in place of nebulous medium \citep[e.g.][]{Snell2002, Brunt2003}. It is with this reasoning that \cite{Blair2008} surveyed galactic molecular clouds for \htwoco{} in an attempt to reconfigure the outer limits of the habitable region and found \htwoco{} observable in molecular clouds at $R_{gal} > 20~\mathrm{kpc}$. CH$_3$OH has also been cited as a potential precursor to larger organic species such as glycoaldehyde \citep{Coutens2015}, acetic acid \citep{Boamah2014}, and ethylene glycol \citep{Zhu2020} while also being deemed a species most likely to survive in benign environments, leading to it being proposed as a viable criterion for defining the GHZ, for which \cite{Bernal2021} found its range of detection to be a compelling $R_{gal} = 12.9$ to $23.5~\mathrm{kpc}$. 

Since its discovery \citep{Snyder1969}, interstellar \htwoco{} has been extensively studied as a biogenic precursor, given its association with numerous biochemically-relevant processes \citep[e.g.]{Ehrenfreund2002, Caselli2012}. Its presence is closely linked to the Galactic Habitable Zone (GHZ) suggesting the potential for life-bearing environments in regions enriched with \htwoco{} \citep{Blair2008}. Protoplanetary disks inherits and preserves \htwoco{}-containing interstellar ice, providing evidence for the presence of prebiotic precursor molecules in planet-forming regions \citep{Evans2025}, and \cite{Punanova2025} most recently confirmed that the formation of \htwoco{} is directly connected to the formation of methanol (CH$_3$OH), which is often referred to as the simplest COM and acts as an abundance reference for other COMs. \cite{Costanzo2007} detailed \htwoco{'s} role in a specific chemomimetic reaction to synthesize components of nucleic acid in the form of formamide (NH$_2$CHO) as an \textit{in situ} precursor, which motivated \cite{Adande2013} to target molecular cloud regions harboring compact hot cores and found widespread NH$_2$CHO distribution in locations ranging from $0.13$ to $9.6~\mathrm{kpc}$ with its greatest possibility of life at a radius of about $2.5~\mathrm{kpc}$.  As such, this necessitates research geared towards constraining the GHZ according to the limits of where \htwoco{} is found. The currently accepted inner and outer limits of the GHZ are about $7$ to $9~\mathrm{kpc}$ from the Galactic Center \citep{Gowanlock2011, Prantzos2007} with our Sun in a serendipitous location of $8~\mathrm{kpc}$.

While \htwoco{} has been scantily detected in the outer Galaxy \citep{Blair2008} and towards the Galactic center \citep{Sandqvist1980}, systematic studies of its distribution in the inner Galaxy are limited. The chemical environment in the neighborhood of the region remains underexplored as sporadic studies discuss separate molecules within limited ranges and are currently insufficient to wholly define the GHZ in terms of its chemical composition. Current habitability models also heavily focus on metallicity and probability to survive annihilating events, but often overlook the role of biogenic precursors in defining these zones. Building on existing studies, by now focusing on the presence of \htwoco{} in the inner Galaxy i.e. $< 7~\mathrm{kpc}$, one of the richest repositories of COMs in the Galaxy \citep[e.g.][]{Zeng2018}, we intend to ascertain to what extent the established inhibitors of habitability affects the synthesis and survival of formaldehyde deeper in the Galaxy, if there are regions in the inner galaxy where prebiotic molecules are shielded from destructive processes, if the presence of significantly abundant formaldehyde in the inner galaxy could speculatively redefine the boundaries of the Galactic Habitable Zone, and mainly to contribute to the usage of prebiotically plausible molecules in honing the contemporary definition of a Galactic Habitable Zone as a whole.

The paper is structured as follows. In Section \ref{sec:method}, we describe the sample selection, observational methods, and data analysis. We present the results of our observation as well as preliminary discussions in Sections \ref{sec:result} and \ref{sec:discussion} respectively. Section \ref{sec:derived} then underlines our procedures in determining the physical parameters of our sample, primarily the galactocentric distances and column densities, while Section \ref{sec:relation} discusses the correlations presenting from said parameters. Finally, in Section \ref{sec:GHZ} we extensively deliberate on the implications of our results on the current understanding of the GHZ bounds. 

\section{Methodology} \label{sec:method}

\subsection{Sample Selection} \label{sec:method:sample}

The molecular cloud sample was drawn from the Bolocam Galactic Plane Survey \citep[BGPS;][]{Aguirre2010}, which mapped a $170~\mathrm{deg}^2$ region of the Galactic plane in the $1.1~\mathrm{mm}$ continuum using the Caltech Submillimeter Observatory (CSO). The survey catalogued 8358 compact sources, with a 98\% completeness level for flux densities between $0.4$ and $60.0~\mathrm{Jy}$ \citep{Rosolowsky2010}. To refine our sample, we selected sources with flux densities in the range $1~\mathrm{Jy} < S_{\mathrm{BGPS}} < 10~\mathrm{Jy}$ and restricted the Galactic latitude to $\|b\| < 1^\circ$. Furthermore, we ensured that each selected source was spatially isolated by requiring the angular separation from its nearest neighbor to exceed the telescope’s beam size (described in the following subsection). This selection yielded a final sample of 215 sources.

\subsection{Observations} \label{sec:method:observations}

The observations of the selected molecular clouds were conducted using the Nanshan 25-m radio telescope\footnote{The telescope has since been upgraded to 26 meters, as of late 2015.}, operated by the Xinjiang Astronomical Observatory in Urumqi, China ($43^{\circ}28'17''\mathrm{N}$, $87^{\circ}10'41''\mathrm{E}$), situated at an altitude of 2080 meters above sea level. The telescope is equipped with a C-band cryogenic receiver with a system temperature of $23~\mathrm{K}$. It offers a beam efficiency of 65\% and achieves a pointing and source tracking accuracy of better than 15 arcseconds. The full-width at half-maximum (FWHM) of the primary beam at the telescope operating frequencies is approximately 10 arcminutes. A diode noise source was employed to calibrate the spectrum at the start of each observation period, with flux calibration uncertainties within 10\%.

The observations were carried out between November 2012 and February 2014 using the Digital Filter Bank (DFB) for spectrometry. The frequency setup was centered at $4851.9102~\mathrm{MHz}$, with a bandwidth of $64~\mathrm{MHz}$ divided into 8192 channels, corresponding to a spectral and velocity resolution of $7.8~\mathrm{kHz}$ and $0.48~\mathrm{km}~\mathrm{s}^{-1}$, respectively. Rest frequencies of $4829.6594~\mathrm{MHz}$ and $4874.1570~\mathrm{MHz}$ were adopted to measure the velocities of \htwoco{} ($\mathrm{K}_a = 1_{10} - 1_{11}$) absorption and \rrl{} emission. Observations were conducted in the ON/OFF mode. Each source was initially observed with a 24-minute integration, with an additional 24-minute integration performed if the signal-to-noise ratio was found to be less than 3. To ensure the system's reliability, the W3 giant molecular cloud was observed as a reference source.

\subsection{Data Reduction} \label{sec:method:reduction}

The spectral data were reduced using the CLASS software, part of the Grenoble Image and Line Data Analysis Software (GILDAS) package. Baseline subtraction using polynomial baselines was applied to all spectra before fitting Gaussian profiles to the detected lines, with negative profiles for \htwoco{} absorption and positive profiles for \rrl{} emission. Gaussian fits were performed for spectral features above $3\sigma$ (peaks above $2.5\sigma$ were also considered on a case-by-case basis).

\subsection{\bh{} Cataloging} \label{sec:method:catalogue}

To further analyze the selected BGPS sources and their results, we associate each with a single \htwoco{} line and a single \rrl{} line, if detected. In cases where multiple features are present, we select the one with the highest brightness temperature. Line fluxes are then derived using the Rayleigh-Jeans approximation, assuming the telescope's beam FWHM represents the angular extent of each source\footnote{Although these fluxes are negative, our analyses consider them as positive values for simplicity, which we refer to as \htwoco{} absorption fluxes.}. We refer to this catalog as the \bh{} catalog, shown in Table \ref{tab:master}.

\section{Results} \label{sec:result}

\begin{figure}
    \centering
    \includegraphics[width=0.65\linewidth]{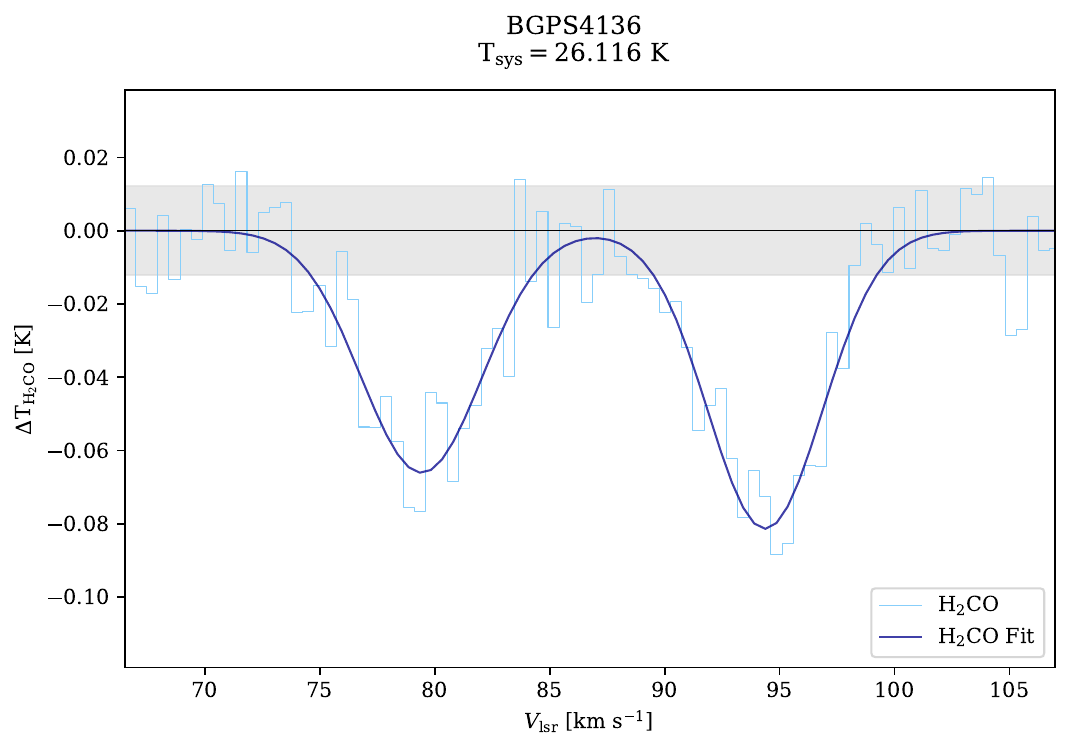}
    \caption{Spectrum of BGPS4136 showing \htwoco{} absorption lines. The grey area denotes the $1\sigma$ baseline noise. Note that this source doesn't possess any \rrl{} lines and therefore is not included in the plot.}
    \label{fig:h2co_spectra}
\end{figure}

\begin{figure}
    \centering
    \includegraphics[width=0.65\linewidth]{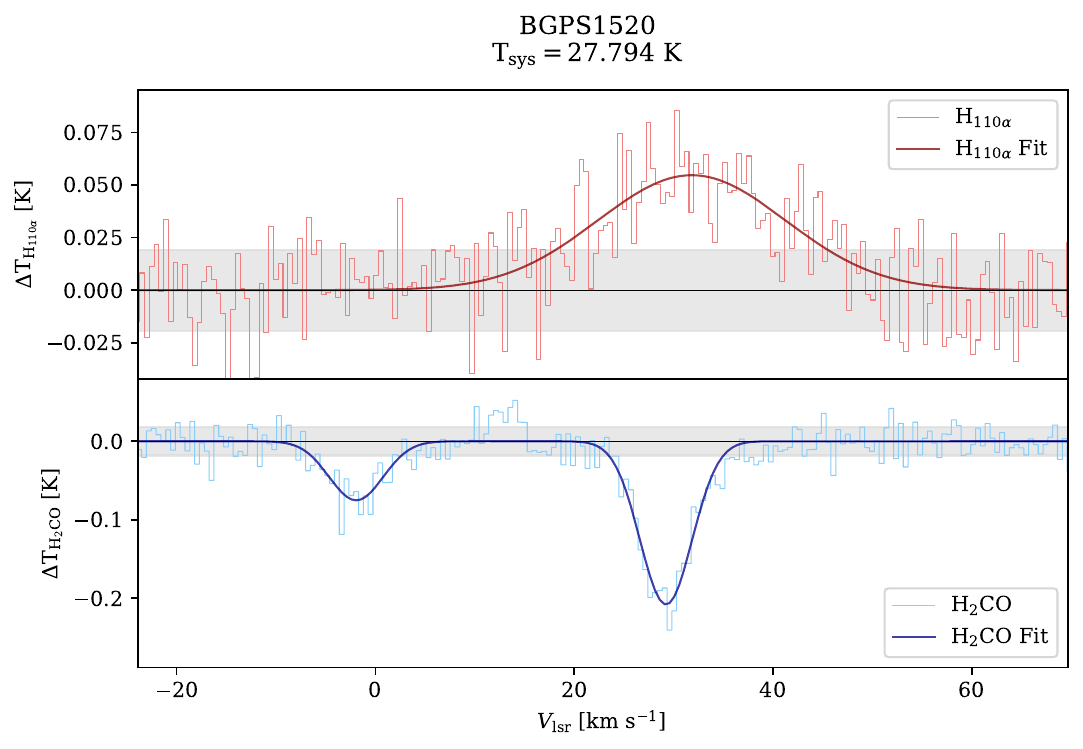}
    \caption{Spectrum of BGPS1520 showing a \rrl{} emission line (in the top panel, and \htwoco{} absorption lines in the bottom panel). Similar to Figure \protect\ref{fig:h2co_spectra}, the gray \highlight{area} denotes the $1\sigma$ baseline noise.}
    \label{fig:rrl_spectra}
\end{figure}

\begin{figure}
    \centering
    \includegraphics[width=0.85\linewidth]{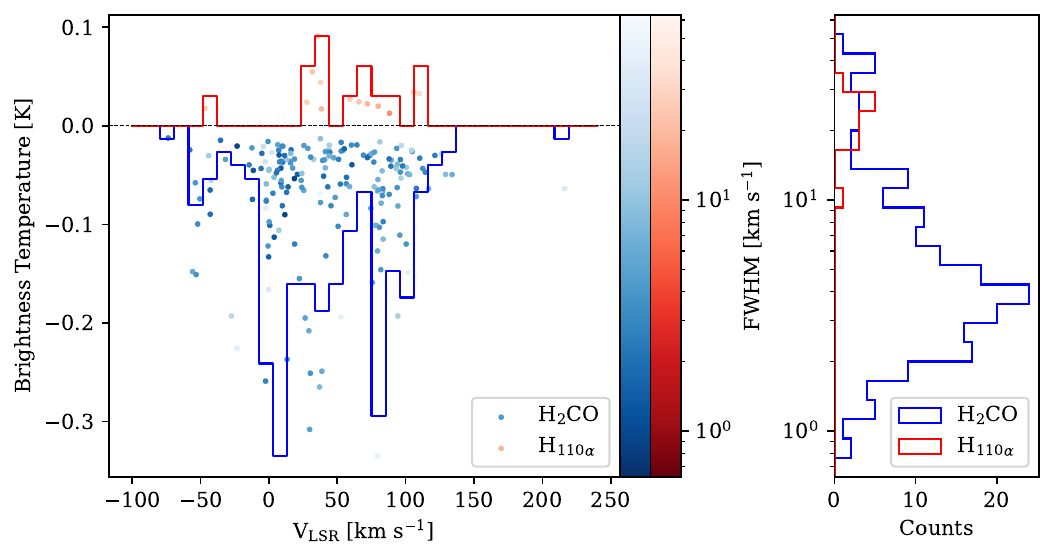}
    \caption{Distribution of parameters for all Gaussian-fitted \htwoco{} and \rrl{} lines. Left: Plot of the lines' brightness temperatures against their central velocities, with colors corresponding to their FWHM. The histograms represent the distribution of the lines' velocities; blue for \htwoco{} and red for \rrl{}. Right: Distribution of the FWHM of the lines.}
    \label{fig:line-distribution}
\end{figure}

A list of all the selected 215 BGPS sources with their \htwoco{} and \rrl{} detection flags are shown in Table \ref{tab:detection}. 88 of the sources were detected with \htwoco{} absorption (40.93\%), and 11 of those sources also possess \rrl{} emission (5.12\%). None of the sources were detected with only \rrl{} lines. Observational parameters for sources with detections are listed in Table \ref{tab:observation}. A total of 183 \htwoco{} absorption lines were detected, with 56 sources (64.64\%) exhibiting more than one spectral feature. An example spectrum of a source displaying \htwoco{} absorption is presented in Figure \ref{fig:h2co_spectra}. Each detected \htwoco{} line is assigned a molecular cloud (MC) number, and the corresponding Gaussian fitting results for all MCs are summarized in Table \ref{tab:h2co_params}. 13 \rrl{} emission lines were identified, with two sources (BGPS2094 and BGPS5910) displaying double-peaked profiles (18.18\%). Figure \ref{fig:rrl_spectra} illustrates a representative spectrum of a source exhibiting both \rrl{} emission and \htwoco{} absorption. Each detected \rrl{} line is assigned an \hii{} region identifier, with their Gaussian fitting results provided in Table \ref{tab:rrl_params}. The statistics of the Gaussian-fitted parameters for both \htwoco{} and \rrl{} lines are presented in Table \ref{tab:statistics}, while Figure \ref{fig:line-distribution} visualizes the distributions of brightness temperatures, velocities, and line widths for all fitted profiles.

\begin{deluxetable*}{ccccccc}
    \tablecaption{Statistics of Gaussian-fitted parameters for \htwoco{} and \rrl{} lines detected.\label{tab:statistics}}
    \tablehead{
        \colhead{Line} & \multicolumn{3}{c}{\htwoco{}} & \multicolumn{3}{c}{\rrl{}} \\
        \colhead{Parameter} & \colhead{Flux} & \colhead{Velocity} & \colhead{FWHM} & \colhead{Flux} & \colhead{Velocity} & \colhead{FWHM} \\
        & \colhead{(K)} & \colhead{($\mathrm{km}~\mathrm{s}^{-1}$)} & \colhead{($\mathrm{km}~\mathrm{s}^{-1}$)} & \colhead{(K)} & \colhead{($\mathrm{km}~\mathrm{s}^{-1}$)} & \colhead{($\mathrm{km}~\mathrm{s}^{-1}$)}
    }
    \startdata
    \multirow{2}{*}{Minimum} & -0.012 & -73.71 & 0.88 & 0.013 & -46.76 & 10.69 \\
    & BGPS2864 MC1 & BGPS2864 MC1 & BGPS5813 MC2 & BGPS5910 \hii{}\,a & BGPS6661 \hii{}\,a & BGPS5910 \hii{}\,b \\[0.5em]
    \multirow{2}{*}{Maximum} & -0.335 & 216.30 & 50.45 & 0.091 & 109.90 & 31.98 \\
    & BGPS0512 MC4 & BGPS0548 MC4 & BGPS0927 MC1 & BGPS2094 \hii{}\,b & BGPS4372 \hii{}\,a & BGPS4372 \hii{}\,a \\[0.5em]
    Mean & -0.069 & 40.56 & 7.00 & 0.032 & 54.30 & 23.07 \\[0.5em]
    Median & -0.050 & 38.77 & 4.12 & 0.024 & 59.28 & 23.31 \\
    \enddata
\end{deluxetable*}

\defcitealias{Guo2016}{G16} To comprehensively compare our results with known detections, we compiled all available literature reporting positive detections of the \htwoco{} ($\mathrm{K}_a = 1_{10} - 1_{11}$) absorption line in the Milky Way to produce an extended \htwoco{} list. \citet[\citetalias{Guo2016} hereafter]{Guo2016} previously cataloged all $4.8~\mathrm{GHz}$ \htwoco{} absorption detections up to March 2014, totaling 2,241 unique sources. Since then, a few additional detections have been reported towards various astrophysical environments, including low- and high-mass star-forming regions \citep{Araya2015, Chen2017, Gong2023}, molecular clouds \citep{Yan2019}, and giant molecular clouds \citep{Ginsburg2015a, Komesh2019, Mahmut2024}. 

\begin{figure}
    \centering
    \includegraphics[width=0.65\linewidth]{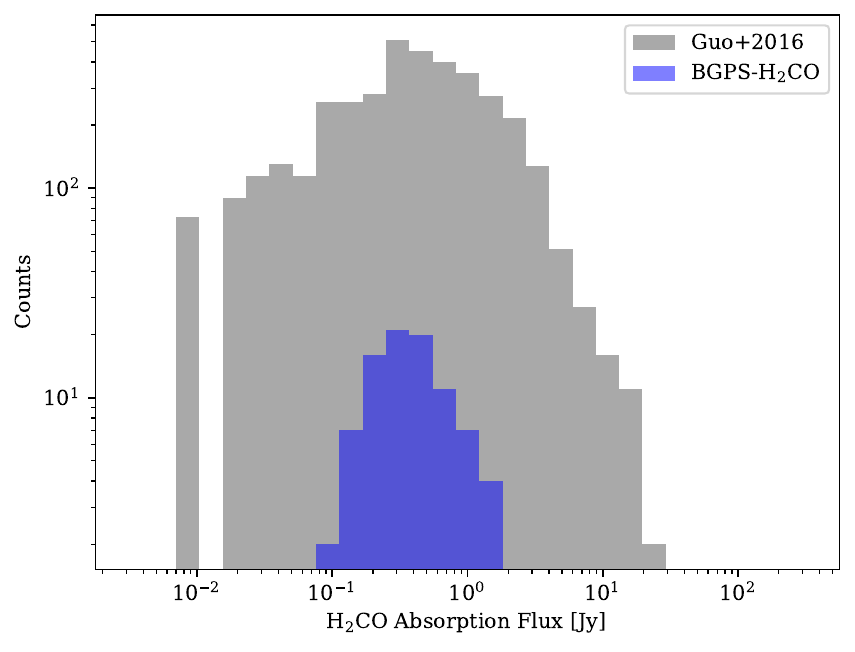}
    \caption{Distribution of \htwoco{} absorption fluxes for \citetalias[labeled Guo+2016]{Guo2016} and our \bh{} cataloged.}
    \label{fig:compare_flux}
\end{figure}

\begin{figure}
    \centering
    \includegraphics[width=0.75\linewidth]{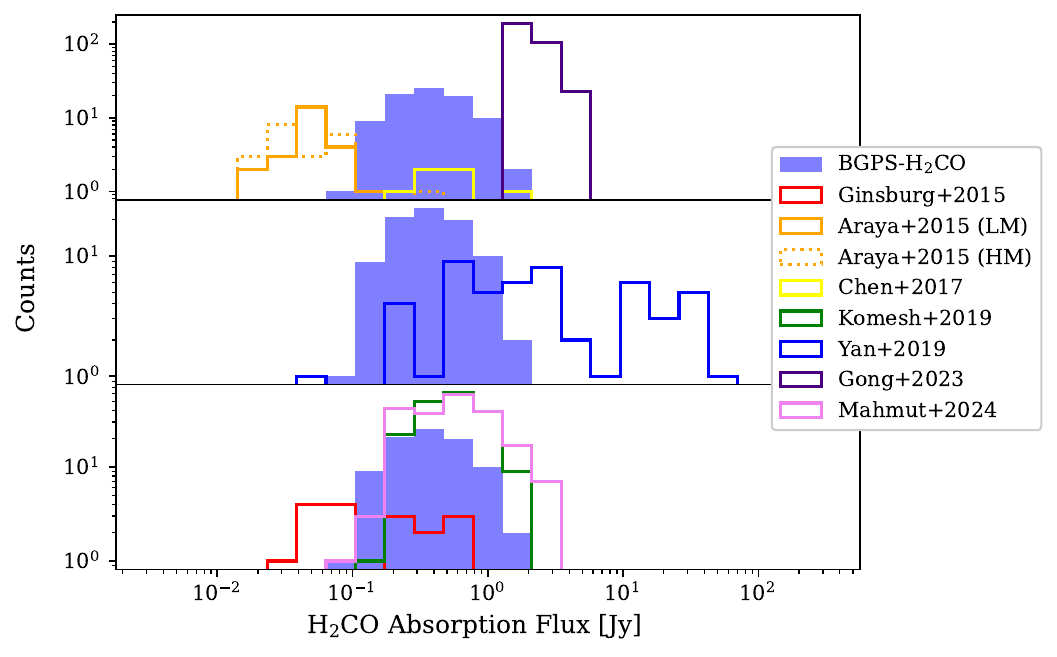}
    \caption{Distribution of \htwoco{} absorption fluxes for literatures after March 2014 and \bh{}. References: \citealt{Ginsburg2015a} (Ginsburg+2015), \citealt{Araya2015} (Araya+2015, LM and HM represent low-mass and high-mass SFRs respectively), \citealt{Chen2017} (Chen+2017), \citealt{Komesh2019} (Komesh+2019), \citealt{Yan2019} (Yan+2019), \citealt{Gong2023} (Gong+2023), and \citealt{Mahmut2024} (Mahmut+2024).}
    \label{fig:compare_flux_other}
\end{figure}

\begin{figure}
    \centering
    \includegraphics[width=0.65\linewidth]{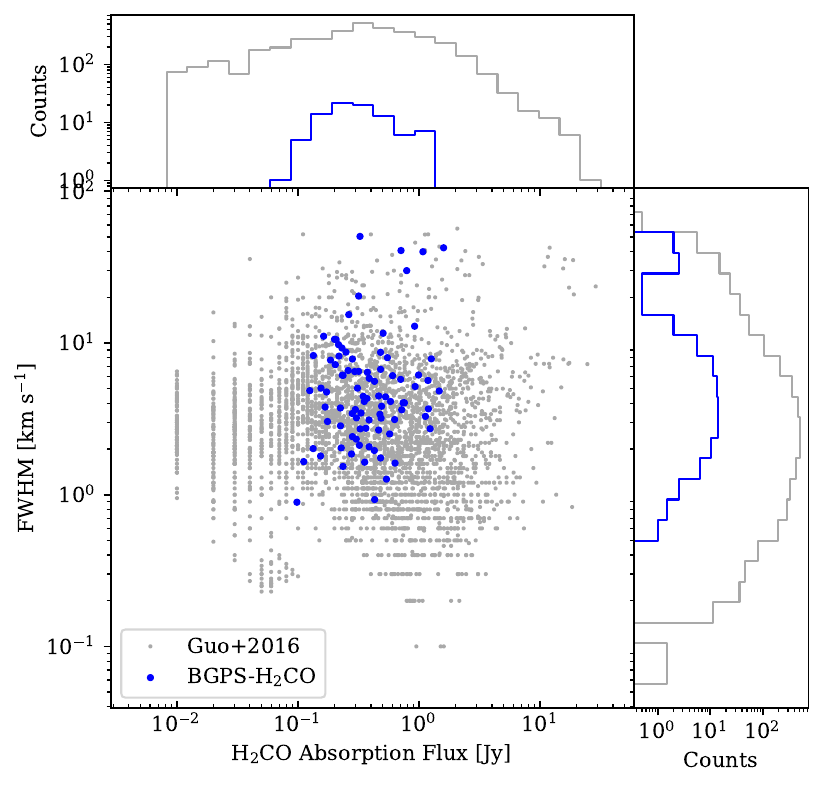}
    \caption{Distribution of \htwoco{} absorption flux and FWHM for \citetalias[labeled Guo+2016]{Guo2016} and \bh{}.}
    \label{fig:compare_flux-fwhm}
\end{figure}

Shown in Figure \ref{fig:compare_flux} and \ref{fig:compare_flux_other} are comparisons of the distributions of \htwoco{} absorption fluxes for \bh{} with \citetalias{Guo2016} and with the other catalogs, respectively. The fluxes in \bh{} falls within the range of \citetalias{Guo2016}, with a mean value within an order of magnitude i.e. $0.463~\mathrm{Jy}$ and $0.869~\mathrm{Jy}$ respectively. Similarly, the flux distribution of \bh{} also aligns well with that of the other catalogs, exhibiting a comparable range. In particular, the results from observations using the Nanshan telescope \citep{Komesh2019, Mahmut2024} closely match those of \bh{}. Observations using telescopes with larger diameters \citetext{e.g. Arecibo Telescope, \citealt{Ginsburg2015a, Araya2015}; Tianma Radio Telescope, \citealt{Yan2019}} were able to detect \htwoco{} at lower fluxes due to their higher sensitivity. Figure \ref{fig:compare_flux-fwhm} shows the distribution of \htwoco{} absorption fluxes and line FWHM for \bh{} compared to \citetalias{Guo2016}, and the line parameters in \bh{} are consistent. After cross-matching the extended \htwoco{} list with our \bh{} catalog, we found that 59 out of the 88 \bh{} sources are newly detected.

\section{Discussion} \label{sec:discussion}

\subsection{Coincident \htwoco{} and \rrl{} Detections} \label{sub:coincident}

Based on our results, there were no detections of \rrl{} without an accompaniment of \htwoco{}, and only 11 (12.5\%) of the sources possessing \htwoco{} absorption also have \rrl{} emission line(s). RRLs and \htwoco{} are not typically detected in the same regions as they usually originate from fundamentally different environments. RRLs trace hot ($\sim 10000~\mathrm{K}$), ionized gas in \hii{} regions, which are created and maintained by UV radiation from massive stars. In contrast, \htwoco{} absorption lines come from cold, dense molecular gas shielded from ultraviolet (UV) radiation. The intense UV radiation in \hii{} regions destroy molecules like \htwoco{}, establishing a sharp boundary between ionized and molecular gas \citep{Hollenbach1999}.

As for the BGPS sources with both detections, multiple explanations are possible. Photodissociation regions (PDRs) in massive SFRs allows for the coexistence of ionized and molecular gases, usually in the boundaries of \hii{} regions \citep{Hollenbach1999}. However, pressure-driven expansion of the \hii{} region can further separate them, leading to a rarefied ionized region and molecular gas dominance in surrounding clouds \citep{Keto2002}. Presence of dense gases in molecular clumps can also shield \htwoco{} from the strong UV radiation, preventing its destruction \citep{Cuadrado2017}. In young and complex massive star-forming regions, localized areas with varying physical conditions can exist in close proximity, particularly when ionized gas has not yet fully dispersed the surrounding molecular cloud \citep[e.g.][]{Bally1980}. Additionally, since massive stars are intrinsically rare according to the stellar mass function, such unique regions are relatively uncommon.

As for the line parameters, no correlation was seen between the \htwoco{} absorption fluxes and \rrl{} emission fluxes. Similarly, the FWHMs of the coincident lines do not exhibit any direct relationship, but \rrl{} lines generally exhibit broader FWHMs than \htwoco{}. The median FWHM of \rrl{} emission is $23.31~\mathrm{km}~\mathrm{s}^{-1}$, consistent with previous observations of \rrl{} in \hii{} regions \citep[e.g.,][]{Quireza2006}. In comparison, the median FWHM of \htwoco{} absorption is $4.12~\mathrm{km}~\mathrm{s}^{-1}$, exceeding the range expected from thermal broadening and hyperfine structure effects \citep[see][]{Yuan2014, Guo2016}. The broader widths of RRLs are primarily attributed to turbulence and large-scale ordered motions within molecular clouds. In young \hii{} regions, the presence of large-scale ionized gas motions around the central protostar also contributes to the line broadening \citep{Sewilo2004, Keto2008}. In contrast, molecular lines, including \htwoco{}, generally have narrower FWHMs, as molecular gas is less spatially extended than \hi{} and \hii{} regions \citep{Menten1996, Koch2019}.

Shown in Figure \ref{fig:bgps_h2co-h110a} is the relation between the \htwoco{} and \rrl{} line velocities in sources with coincident detections. The \htwoco{} velocities correlate close to a $1:1$ relation with the \rrl{} velocities with a high correlation coefficient, suggesting that \htwoco{} and \rrl{} traces the same physical properties of each molecular cloud \citep{Pabst2024, Khan2024}. 

\begin{figure}
    \centering
    \includegraphics[width=0.65\linewidth]{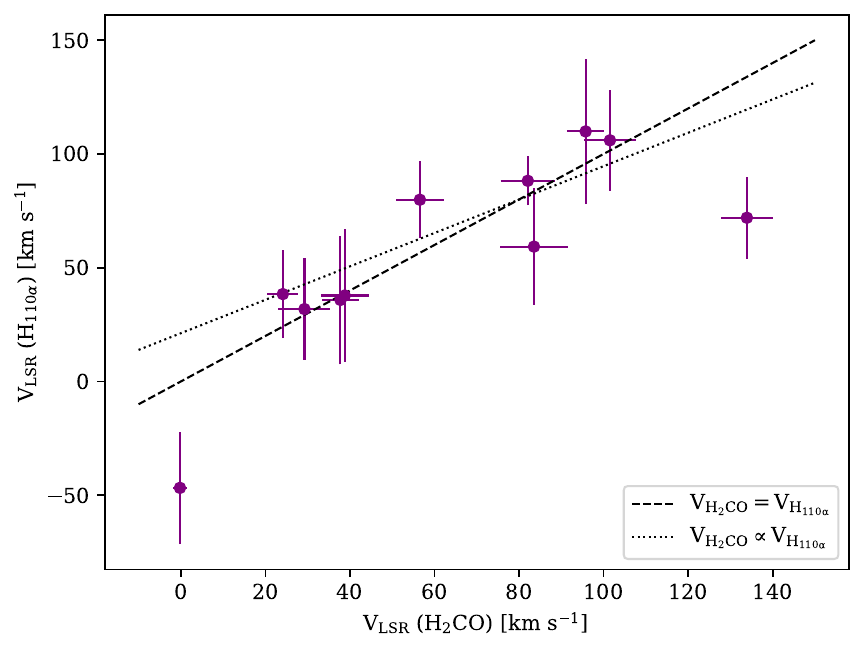}
    \caption{Comparison between \htwoco{} and \rrl{} line velocities in \bh{}. The dashed line represents a exact velocity correlation, while the dotted line represents a best-fit linear relation: $V_{\mathrm{LSR}}~(\mathrm{H_{110\alpha}}) = (0.73\pm0.18) \times V_{\mathrm{LSR}}~(\mathrm{H_2CO}) + (21.2\pm12.7)~\mathrm{km}~\mathrm{s}^{-1}$. The Pearson's correlation coefficient for this fit is $0.8030$.}
    \label{fig:h2co-h110a}
\end{figure}

\subsection{Millimeter and \htwoco{} Absorption Flux} \label{sub:discussion:mm-flux}

\begin{figure}
    \centering
    \begin{minipage}{0.48\textwidth}
        \includegraphics[width=0.95\linewidth]{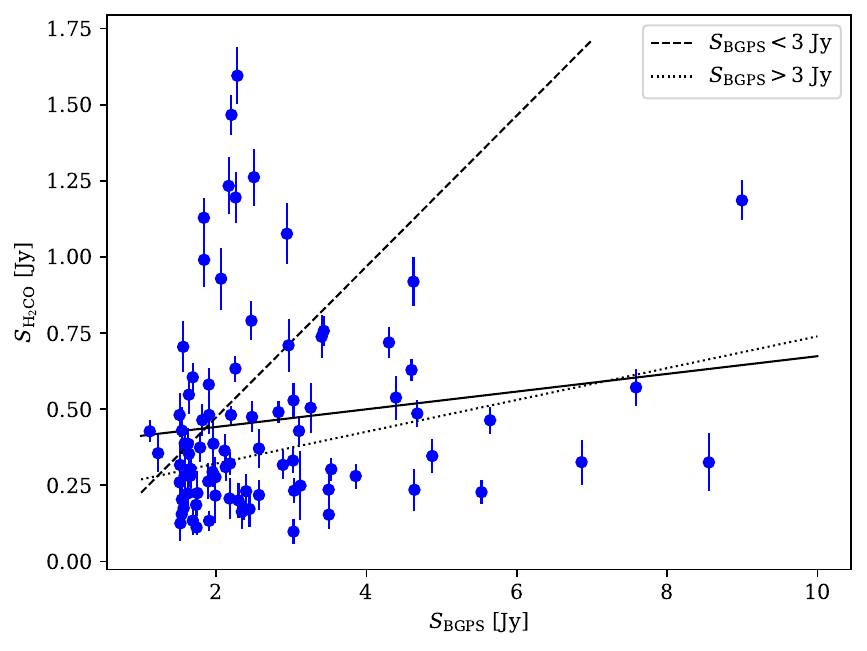}
    \end{minipage}%
    \begin{minipage}{0.48\textwidth}
        \includegraphics[width=0.95\linewidth]{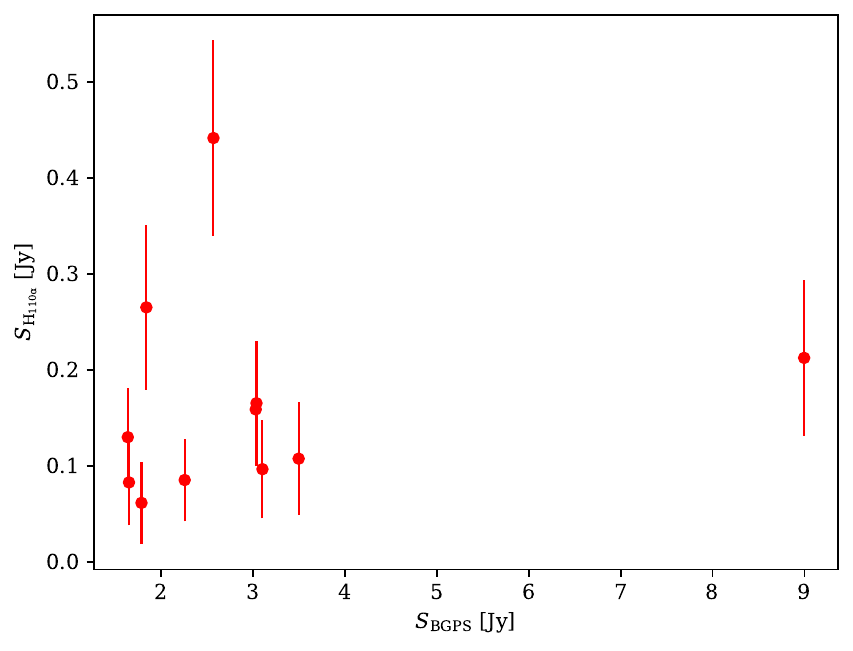}
    \end{minipage}
    \caption{Comparison between BGPS fluxes with \htwoco{} and \rrl{} fluxes in \bh{}. Left: Plot of \htwoco{} against BGPS $1.1~\mathrm{mm}$ fluxes. The solid line represents the best-fit linear relation: $S_{\mathrm{\htwoco{}}} = (0.03\pm0.02) \times S_{\mathrm{BGPS}} + (0.38\pm0.07)~\mathrm{Jy}$. The dashed and dotted lines show separate linear fits for sources below and above the $3~\mathrm{Jy}$ threshold, given by $S_{\mathrm{\htwoco{}}} = (0.25\pm0.10) \times S_{\mathrm{BGPS}} + (-0.02\pm0.21)~\mathrm{Jy}$ and $S_{\mathrm{\htwoco{}}} = (0.05\pm0.03) \times S_{\mathrm{BGPS}} + (0.22\pm0.14)~\mathrm{Jy}$, respectively. The Pearson's correlation coefficients for all three relations are $0.1376$, $0.2968$, and $0.3528$ respectively. Right: Plot of \rrl{} against BGPS $1.1~\mathrm{mm}$ fluxes. The Pearson's correlation coefficient for this relation is $0.1556$.}
    \label{fig:bgps_h2co-h110a}
\end{figure}

Shown in Figure \ref{fig:bgps_h2co-h110a} is the comparison of the \htwoco{} absorption and \rrl{} emission fluxes with the $1.1~\mathrm{mm}$ BGPS fluxes. The relationship between \htwoco{} flux and $1.1~\mathrm{mm}$ continuum flux in \bh{} exhibits significant dispersion, with an overall very weak correlation ($r = 0.1376$). However, when separating the catalog at a $3~\mathrm{Jy}$ threshold, a moderate correlation is present in both the low and high flux regimes ($r = 0.2968$ and $r = 0.3528$ respectively)\footnote{The flux threshold is obtained by testing over all possible flux levels and maximizing both correlation coefficients.}. Each section has different slopes, and this bears resemblance to the results of \cite{Tang2014} who found that \htwoco{} absorption correlates with $6~\mathrm{cm}$ continuum at low flux levels, but eventually becomes more scattered at high fluxes, likely due to changing excitation conditions. However, while their study focused on radio continuum, our work examines the relationship between \htwoco{} flux and $1.1~\mathrm{mm}$ far-infrared continuum, suggesting that different physical mechanisms may be involved. 

Several previous studies have investigated the link between \htwoco{} absorption and infrared flux at shorter wavelengths, generally finding weak or moderate correlations as well. \cite{Du2011} identified two branches in the relation between \htwoco{} and $100~\mathrm{\mu m}$ infrared flux: A strong correlation for sources with \rrl{} detections, and a weaker, more dispersed correlation for sources with only \htwoco{} absorption. \cite{Yuan2014} also found a similar trend but with sources without \rrl{} emission exhibiting no significant dependence. \cite{Guo2016} noted that higher-resolution observations significantly improved the correlation strength, emphasizing the role of beam dilution in these observations. Our results, which show a moderate correlation when split into two categories, are consistent with these trends. This suggests that while the far-infrared continuum may contribute to \htwoco{} excitation, additional effects e.g. local radiation fields and gas density variations likely play a role in the observed dispersion.

One possible interpretation is that at low far-infrared fluxes, the \htwoco{} absorption is more dependent on the cosmic microwave background (CMB), resulting in a moderately strong correlation \citep[e.g.][]{Liszt2016}. \cite{Townes1969} first proposed sub-CMB cooling mechanism of \htwoco{} that allows for effective absorption at low temperatures, and since been confirmed by various observations \citep[e.g.][]{Troscompt2009, Darling2012}. However, as the continuum flux increases, additional effects such as optical depth variations, local heating, and changes in \htwoco{} column density introduces more complexity. Various models have suggested that strong background continuum emission can heat up \htwoco{} and modify its excitation, potentially leading to partial thermalization of the absorption line \citep{Tang2014, Gerin2024, Bu2024}. In \highlight{more} extreme cases, \htwoco{} maser emission has been detected from multiple high-mass SFRs \citep[e.g.][]{Araya2008, Ginsburg2015b}. Observations and simulations have also shown that \htwoco{} excitation temperature can increase above the CMB when electron fractions are high \citep{Turner1993, Gerin2024}. These stacking dependence results in the breakdown in the correlation at higher fluxes, and consideration of various non-local thermodynamic equilibrium (non-LTE) effects may help clarify the anomaly. 

\section{Derived Parameters} \label{sec:derived}

To study our \htwoco{} results in the context of GHZ, we derive two parameters: The galactic location of our catalogue, and the physical parameters of the \htwoco{} detection. 

\subsection{Kinematic and Galactocentric Distance Cross-Matching} \label{sec:derived:kd-cross}
\defcitealias{MivilleDeschnes2017}{MD17}

The first step in examining the galactic location dependence of \bh{} properties is to determine each source's galactocentric distance i.e. its distance to the \highlight{Galactic Center}. The most accurate method for this involves parallax measurements of star-forming regions using Very Long Baseline Interferometry \citep[VLBI; see][]{Reid2014}, though this approach is resource-intensive. Alternatively, kinematic distances (KD) can be estimated using the source's $V_{\mathrm{LSR}}$ and a Galactic rotation model, either through analytical solutions \citep{Anderson2012} or Monte Carlo methods \citep{Wenger2018}. However, in the inner Galaxy, the kinematic distance ambiguity (KDA) introduces additional challenges due to the two possible solutions for KD. This ambiguity can often be resolved using \hi{} self-absorption \citep{Kolpak2003, RomanDuval2009} or Bayesian inference techniques \citep{Reid2019}.

However, the limited number of \rrl{} detections in our \bh{} sample precludes statistically robust analysis for their KD. Additionally, the usage of \htwoco{} absorption lines for KD and KDA resolution has proven to be somewhat unreliable, primarily due to the lower filling factor of molecular gas compared to atomic gas \citep[see][]{Kolpak2003}. 

To overcome these challenges, we adopt the molecular cloud catalogue compiled by \citet[\citetalias{MivilleDeschnes2017} hereafter]{MivilleDeschnes2017}, which is based on the Milky Way $^{12}$CO 1-0 survey conducted by \cite{Dame2001}. The authors identified molecular clouds within $|b| < 5^{\circ}$ by applying a hierarchical clustering algorithm and Gaussian decomposition to the CO maps. The kinematic distance of each cloud was then derived using the standard $V_{\mathrm{LSR}}$ method \citep{RomanDuval2009}, while the KDA is resolved using a modified version of Larson’s scaling relation, incorporating surface density effects i.e. $\sigma_{v} \propto (R\Sigma)^{0.43}$ \citep{Heyer2009}.

To assign KDs and galactocentric distances to our \bh{} sources, we cross-matched the \bh{} catalogue (including non-detections) with the \citetalias{MivilleDeschnes2017} dataset, using a matching threshold based on our instrument's beam FWHM. This resulted in successful cross-matching of 164 out of 215 sources, including 70 with detected \htwoco{} emission. Notably, all 11 sources exhibiting \rrl{} lines were also successfully matched. The results of the cross-matching is appended within the master \bh{} catalogue in Table \ref{tab:master}.

\subsection{\htwoco{} Column Density Calculations} \label{sec:derived:column_density}

To further quantify the physical properties of the sources in \bh{}, we derive their column densities following a similar methodology to \citet{Mahmut2024}. The apparent optical depth, $\tau_{\mathrm{app}}$, is estimated using standard radiative transfer principles:
\begin{equation}
\tau_{\mathrm{app}} = - \ln{\left(1 + \frac{T}{T_{\mathrm{c}} + T_{\mathrm{cmb}} - T_{\mathrm{ex}}}\right)}
\end{equation}
where $T$, $T_{\mathrm{c}}$, $T_{\mathrm{cmb}} = 2.725~\mathrm{K}$, and $T_{\mathrm{ex}}$ denote the line brightness temperature, continuum temperature, CMB temperature, and excitation temperature, respectively. For simplicity, we assume the sources lack strong $6~\mathrm{cm}$ continuum emission and adopt an excitation temperature of $1~\mathrm{K}$, and then apply the approximation $\tau \approx \tau_{\mathrm{app}}$ \citep{Ginsburg2015a}. Using this optical depth, the \htwoco{} column density for each \bh{} source is then determined following \citet{Mangum2015, Mahmut2024}:
\begin{equation}
N_{\rm\htwoco{}} = 7.3\times10^{23}~\mathrm{cm}^{-2} \times \int \tau~\mathrm{d}\nu
\end{equation}
where $\int\mathrm{d}\nu$ represents the FWHM of the detected \htwoco{} line. The derived parameters are appended within the master \bh{} catalogue in Table \ref{tab:master}.

\section{\htwoco{} Relation with Galactic Location} \label{sec:relation}

\begin{figure}
    \centering
    \begin{minipage}{0.48\textwidth}
        \includegraphics[width=0.95\linewidth]{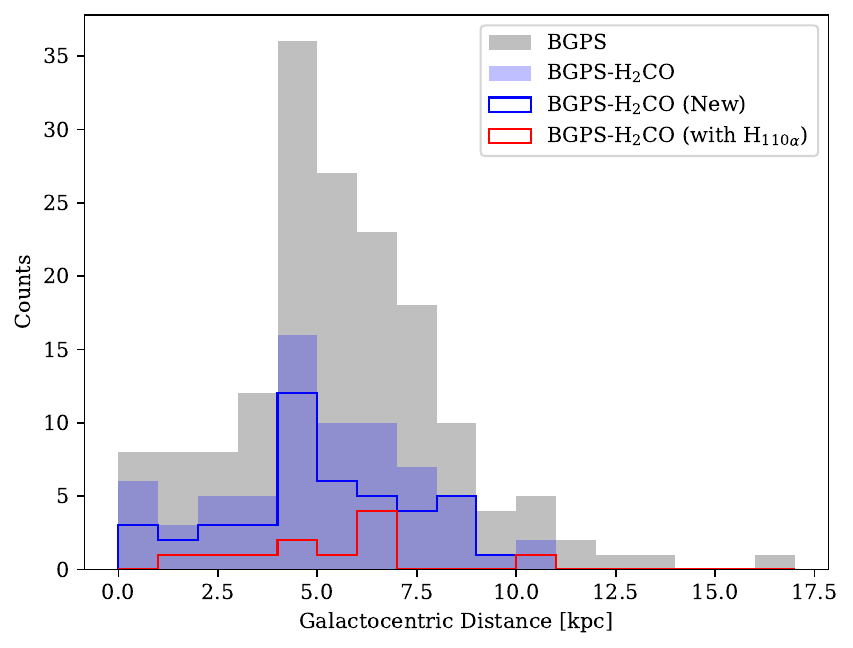}
    \end{minipage}%
    \begin{minipage}{0.48\textwidth}
        \includegraphics[width=0.95\linewidth]{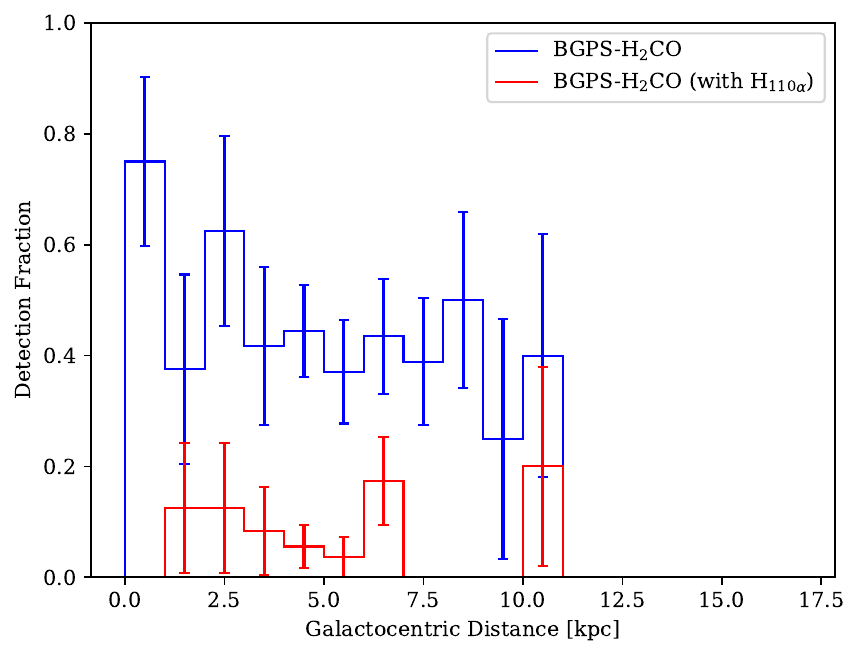}
    \end{minipage}
    \caption{Results of cross-matching between our BGPS sample and the \citetalias{MivilleDeschnes2017} catalog. Right: Distribution of galactocentric distance for all the cross-matched BGPS sources in our sample. Left: Plot of \htwoco{} and \rrl{} detection fraction against galactocentric distance for all the cross-matched BGPS sources in our sample. The Pearson's correlation coefficients for the \htwoco{} and \rrl{} detection fraction against galactocentric distance are -0.8822 and -0.4192, respectively.}
    \label{fig:rgal_distribution}
\end{figure}

164 out of all of the 215 sources in the master \bh{} catalogue were successfully cross-matched with an \citetalias{MivilleDeschnes2017} counterpart, with 70 of them exhibiting \htwoco{} absorption, including all 11 sources that also show \rrl{} emission. The closest source to the Galactic center, BGPS0523 is located at a galactocentric radius of $0.216~\mathrm{kpc}$ and exhibits \htwoco{} absorption but no \rrl{} emission. In contrast, the most distant \htwoco{} detection, BGPS7120 is found to be at $10.769~\mathrm{kpc}$, while BGPS5703 is the farthest cross-matched source from the Galactic center at $16.270~\mathrm{kpc}$, but with no detected \htwoco{} or \rrl{} emission. For completeness, we also report that the closest source to the Galactic center with non-detections of both lines is BGPS0352 at $0.219~\mathrm{kpc}$. Figure \ref{fig:rgal_distribution} describes the \htwoco{} and \rrl{} detection fraction with respect to their galactocentric distance. 

\begin{figure}
    \centering
    \includegraphics[width=0.65\linewidth]{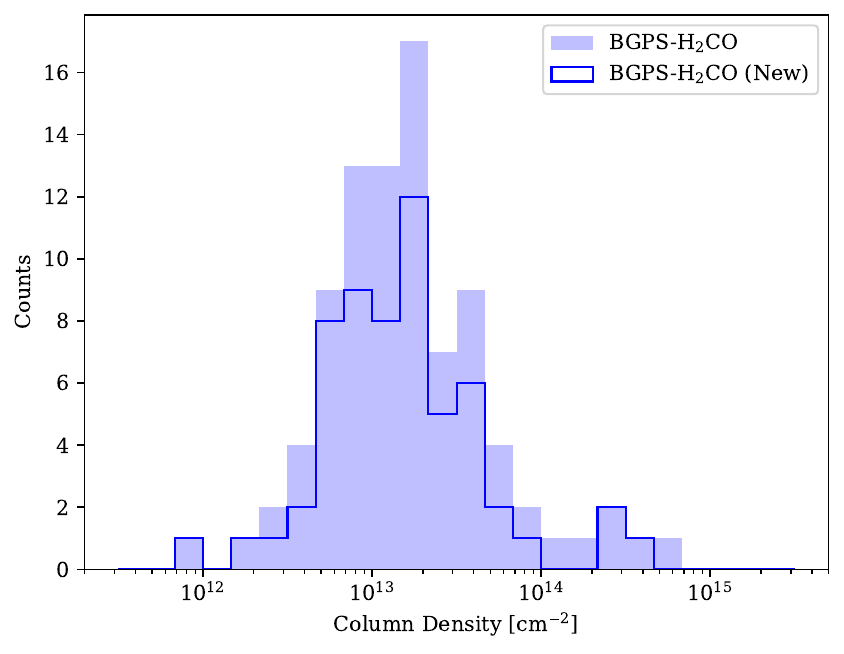}
    \caption{Distribution of calculated column densities from our \bh{} catalog.}
    \label{fig:col-dens_distribution}
\end{figure}

Figure \ref{fig:col-dens_distribution} shows the distribution of \htwoco{} column densities for our \bh{} sample. The \htwoco{} column densities obtained for \bh{} sources are within the range of $7.82\times10^{11}$ to $6.69\times10^{14}~\mathrm{cm}^{-2}$, with a median value of $1.51\times10^{13}~\mathrm{cm}^{-2}$. These values are consistent with those reported in molecular cloud surveys \citep[e.g.][]{Yan2019} and targeted molecular cloud observations \citep{Gong2023, Mahmut2024}, even across different transitions \citep[e.g.][]{Zhao2024}. For massive star-forming regions in particular, the column densities agree with those of high-mass starless cores, but are approximately $2-3~\mathrm{dex}$ lower than more evolved stages such as high-mass protostellar objects and ultra-compact \hii{} regions \citep{Zahorecz2021}. This supports the notion that the \bh{} objects are indeed in early stages of star formation.

\begin{figure}
    \centering
    \includegraphics[width=0.75\linewidth]{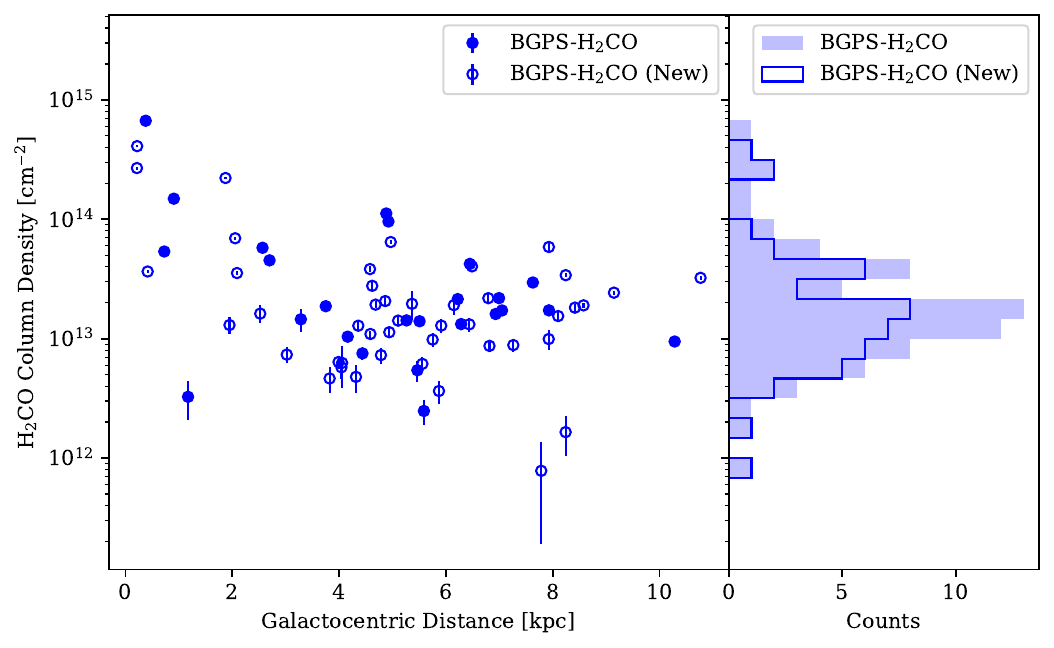}
    \caption{Relation between \bh{} galactocentric distances and their column densities. Left panel: Plot of \bh{} sources' column densities against galactocentric distances. The Pearson's correlation coefficient for this relation is $-0.4622$. Right panel: Distribution of the calculated column densities of the \bh{} cataloged for sources with cross-match counterparts in \citetalias{MivilleDeschnes2017}.}
    \label{fig:col-dens_rgal}
\end{figure}

Consequently, 70 \bh{} sources with associated galactocentric distances also have calculated column densities. Shown in Figure \ref{fig:col-dens_rgal} is the relation between \bh{} galactocentric distances and their column densities. The \htwoco{} detection fractions (see Figure \ref{fig:rgal_distribution}) and column densities both inversely-correlate with galactocentric distance. As \htwoco{} commonly traces star formation, this indicates that such activities are more prominent closer to the Galactic center, distancing the notion that star formation is suppressed in such extreme environments. 

One possible interpretation for the correlations is the presence of common fueling mechanisms for both active galactic nuclei (AGN) and star formation. Tidal torques induced by mergers of gas-rich galaxies i.e. \textit{wet mergers} can drive rapid inflows of gas within galaxies \citep{Hopkins2008}. Similarly, close interactions between galaxies of comparable masses can channel molecular gases towards the center of these galaxies \citep{Koulouridis2006}. Secular processes such as the formation and/or instability of galactic bars and bulges can also facilitate gas inflow, especially in Seyfert galaxies and low-luminosity quasars \citep{Hopkins2008}. In post-starburst galaxies, supernovae and black hole activity can disperse gas and cause inhomogeneity, albeit briefly \citep{Hopkins2008}. All these phenomena contribute to the movement of gas towards the galactic center, which can then trigger starbursts as well as feed AGN accretion, and this has been verified through numerical simulations \citep{Mihos1996, DiMatteo2005}.

AGNs have also been proposed to provide positive feedback by enhancing star formation in galaxies. Models and simulations have demonstrated that outflows from AGNs can over-compress cold gases to form stars, especially for galaxies in their gas-rich phase \citep{Zubovas2013}, for which its effect is dependent on the central AGN properties \citep{ClavijoBohorquez2024}. Observational evidence supporting this scenario includes the detection of polycyclic aromatic hydrocarbons (PAHs) across various spatial scales in AGN-dominated galaxies \citep[e.g.,][]{Sales2013, Zakamska2015, GarciaBernete2021}. Quantitatively, black hole accretion rates have been found to correlate with star formation rates when traced using forbidden molecular lines, especially in a galaxy's inner regions \citep{DiamondStanic2012, EsparzaArredondo2018, MartinezParedes2019}. Although the effect is more pronounced at higher redshifts \citep[$z > 3$;][]{Pouliasis2022}, AGN activity does not universally suppress star formation in nearby galaxies \citep{Mulcahey2022}. 

\section{GHZ Implications} \label{sec:GHZ}

In reconciling the presence of \htwoco{} near the hostile Galactic center despite its significance as a biogenic molecule, this paper draws several inferences. \htwoco{'s} high rotational constants \citep{DeFrees1986}, strong dipole moment \citep{Johnson1972}, and temperature-dependence of the hyperfine coupling constants \citep{Ellinger1980} make it resist degradation, and furthermore persist, notably in dense molecular clouds that provide shielding effects. \htwoco{'s} biogenic role is also more pivotal in the infancy of COMs, serving as a stepping stone rather than a direct contributor to the emergence or sustenance of life, and many of the broader chemical pathways that occur in molecular clouds may be decoupled from the specific conditions required to sustain life. \cite{Araya2014, Molpeceres2021} and \cite{Herbst2022}, in the course of exploring the catalytic processes driving the emergence of complex organic molecules in molecular clouds, extemporaneously indicate the intermediary but fleeting nature of \htwoco{}. Gas-phase and grain-surface processes involving \htwoco{} in the preliminaries go on to synthesize larger molecules \citep{Watanabe2002, Chuang2015, Bergner2017, Carder2021}, including through a three-body mechanism refined by \cite{Jin2020} based on \cite{Fedoseev2015, Fedoseev2017} with \cite{Theule2013} in which \htwoco{} acts a reactive intermediate, either reacting with hydrogen atoms to form the hydroxymethyl radical (CH$_2$OH), which further reacts to produce methanol
\begin{gather}
    \rm H_2CO + H \to CH_2OH \\
    \rm CH_2OH + H \to CH_3OH
\end{gather}
or reacting with the formyl radical (HCO) to form glycoaldehyde, a simple sugar precursor which will then undergo further hydrogenation to form ethylene glycol
\begin{gather}
    \rm H_2CO + HCO \to CH_2OHCHO \\
    \rm CH_2OHCHO + H \to CH_2OHCH_2OH
\end{gather}
all of which present enhance efficiency through the nondiffusive three-body processes, occurring at the more challenging proximities of the Galactic center, rather than its outwards arms. \cite{Evans2025} have also discussed the coupling of \htwoco{} and CH$_3$OH in protoplanetary disks, further lending credence in expositing \htwoco{} near the Galactic center might originate from interstellar ice that has been shielded within dense molecular clouds, allowing it to persist despite the hostile conditions.

\begin{figure*}
    \centering
    \includegraphics[width=0.95\linewidth]{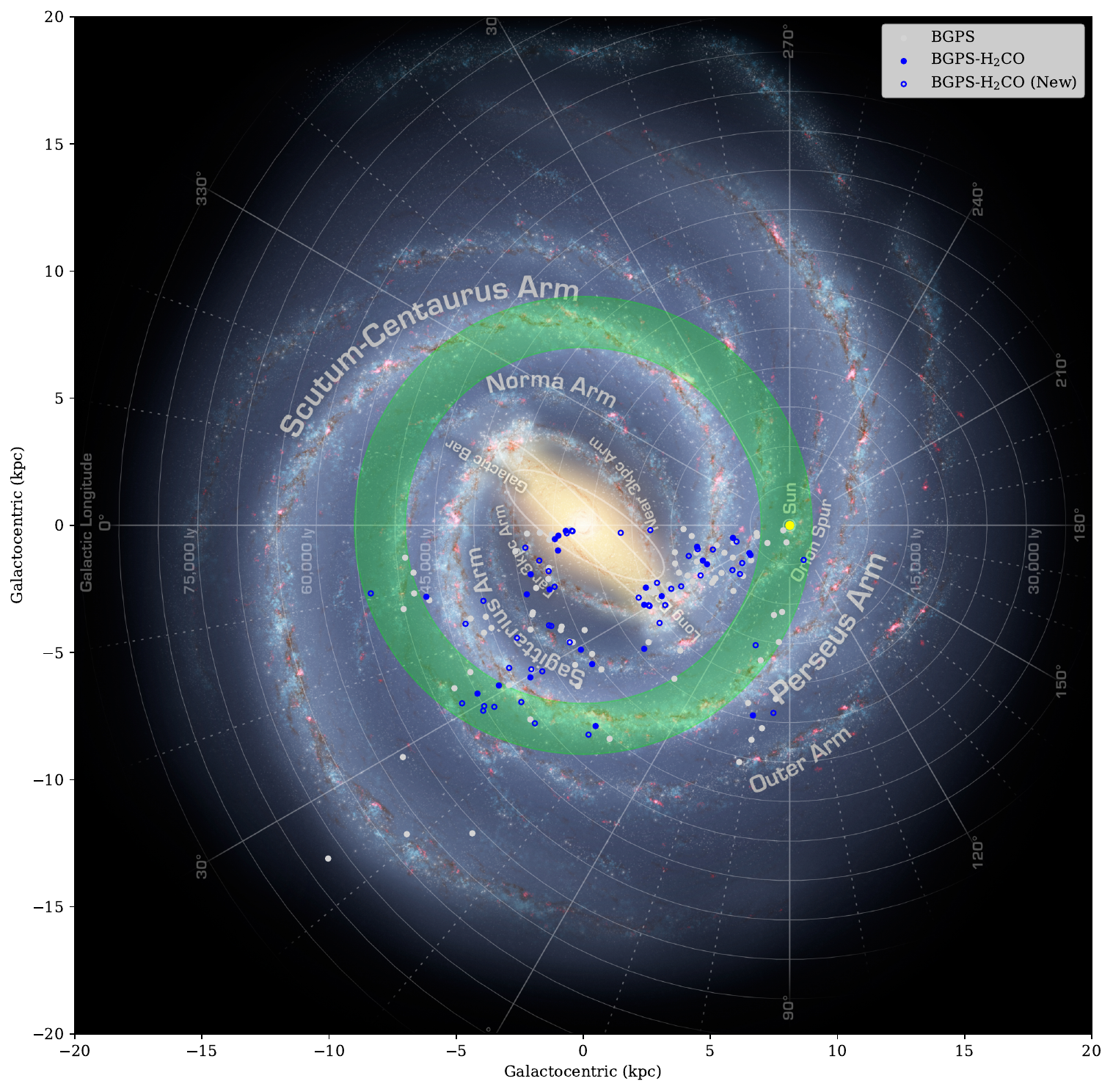}
    \caption{Location of \bh{} sources in the Milky Way based on cross-matching with \citetalias{MivilleDeschnes2017}. The yellow spot denotes the location of the Sun, while the green annuli shows the traditional GHZ between $7$ and $9~\mathrm{kpc}$ from the Galactic Center. Credit (Milky Way Background): NASA/JPL-Caltech/R. Hurt (SSC/Caltech).}
    \label{fig:mw-faceon}
\end{figure*}

While \htwoco{} has been established as an essential biogenic molecule, its widespread distribution in regions across a wide range of galactocentric distances \citetext{mainly \citealt{Blair2008}, also see references in Section \ref{sec:result}} complicates its role as a reliable sole tracer of habitability. Our results show a significant amount of \htwoco{} detections in the inner Galaxy as in the outer Galaxy priorly documented, as near as $0.2~\mathrm{kpc}$ from the Galactic Center. Shown in Figure \ref{fig:mw-faceon} is the location of the cross-matched \bh{} sources in the Milky Way. However, the higher levels of ionizing radiation that can photolyse complex molecules \citep{Morris1996}, higher frequency of supernovae \citep{BlandHawthorn2003}, and higher flux of cosmic rays \citep{Strong2007} in the environment around the Galactic Center have devastating effects on potential life forms. Increased stellar collisional rates \citep{Rose2023} and gravity-induced disruptions \citep{Genzel2010} would also make for an unstable environment for the continuous development of any lifeform that might rise. 

As such, with all considerations taken into account, the sole usage of \htwoco{} as a chemical tracer for GHZ should be tentatively approached. \cite{Fontani2022} demonstrated the ubiquitous distribution of organic molecules across diverse galactic environments and argued how this underscores the need for a more rigorous examination of chemical abundance implying habitability. It would benefit significantly from ancillary observations of other key biomolecules, such as ammonia (NH$_3$), that can react with \htwoco{} and hydrogen cyanide (HCN) to play a critical role in prebiotic chemistry. For instance, NH$_3$ can react with \htwoco{} to produce formamide (HCONH$_2$), a precursor to nucleobases and genetic species \citep{Saladino2012}, while HCN is a fundamental building block for the synthesis of amino acids \citep[via Strecker synthesis;][]{Strecker1850, Kitadai2018} and nucleotides \citep[via polymerization;][]{Oro1961}. Additionally, the co-detection of \htwoco{} and HCN in stable environments, such as protoplanetary disks \citep{Oberg2010} could indicate regions conducive to prebiotic chemistry, including the formation of sugars via the formose reaction \citep{Weber2005}. 

Beyond these molecules, the overall presence of NCHOPS elements \footnote{Nitrogen, Carbon, Hydrogen, Oxygen, Phosphorus, and Sulfur; the six key elements in prebiotic chemistry of the Universe, with earliest mentions by \cite{Pasek2005, Prantzos2007, Pasek2008a}} are also prebiotically significant. These elements are the building blocks of complex biomolecules, including proteins, RNA/DNA, and lipids, which are necessary for the emergence of life \citep{Wald1964, Baross2005, Freire2022}. Phosphorus is a critical component of nucleotides and phospholipids \citep{Pasek2007, Pasek2008b}, and the discovery of phosphorus bearing molecules such as phosphorus monoxide (PO) and phosphorus mononitride (PN) in the outer Galaxy \citep[WB86-621 at $22.6~\mathrm{kpc}$ from the Galactic Center;][]{Koelemay2023} challenges the notion of barrenness of said region for the formation of life. Another essential element for formation of amino acids e.g. cysteine and methionine \citep{Valle2021} is sulfur, which was also observed in a couple star-forming regions \citep{Minh2016}, and the absence of sulfur dioxide (SO$_2$) can be utilized to distinguish exoplanets with hospitable planetary atmospheres and presence of water, further constraining the possibility of life \citep{Jordan2025}\footnote{The absence signifies that sulfur is actually contained within the atmosphere in a evaporation-deposition cycle, allowing for its participation in biological reactions.}. Thus, observations of other chemical tracers, particularly NH$_3$ and HCN, and generally NCHOPS molecules, can further constraint the chemical GHZ in context of the formation of biological prerequisites to the formation of organisms. However, their interpretation must be coupled with the physical constraints \citep[e.g. supernovae frequency, cosmic rays, and stellar density;][]{Lineweaver2004, Gowanlock2011} which can disrupt the stability of prebiotic environments and ultimately affect habitability.

Alternatively, the significance of a biogenic precursor being abundantly proximate to the Galactic Center can be expounded further by leveraging the concept of Galactic Habitable Orbit (GHO) proposed by \cite{Baba2024}. The authors offered the theory of a traveling Sun where the dynamic influence of the Galactic bar and spiral arms had the Sun migrating a radial distance of $3.2$ to $4~\mathrm{kpc}$ from its birth radius near the Galactic Center to its present location. This was previously modeled by \cite{Sellwood2002} whom showed how stars can migrate radially in the Galactic disk through regions with different physical and chemical conditions.  

\highlight{In such manner, to put in the context of this research, a future life-seeding star could traverse different regions of the Galaxy, potentially acquire/store prebiotic molecules and settle in a less-perturbed areas for the essential reactions for an early Earth to occur and not necessarily flourish in the Galactic Center itself. This acquisition, however, rarely occurs directly. Several plausible ways could be during the brief protoplanetary disk phase \citep{Qi2013a, Qi2013b, Ilee2021} or while transiting the dense cores of giant molecular clouds \citep[GMCs; e.g.][]{WidicusWeaver2012}. Another possible pathway involves the indirect delivery of biogenic molecules via intermediary bodies such as comets which can preserve material in their ice reservoirs \citep{Biver2019}. \cite{Milam2006} observed \htwoco{} in comets C/1995 O1 (Hale-Bopp), C/2002 T7 (LINEAR), and C/2001 Q4 (NEAT) and postulated organics like \htwoco{} may detach from the main cometary bodies and seeded planets or enriched environments that contain planetary systems encountered during their flight, one such mechanism being through photodissociation \citep{Feldman2015}.}

The fact that \htwoco{} was previously detected at distances as far as $23~\mathrm{kpc}$ should not imply habitability as well, as it could be the relics of past dynamical processes such as galactic mergers, supernova-driven outflows or the aforementioned stellar migrations which might have redistributed prebiotic molecules across the galaxy. \cite{Minchev2010} demonstrated that stars migrate radially through resonant scattering and would naturally transport with them the chemical composition they inherited at their birth location, including molecules like \htwoco{}. \cite{Ibata1994, Ibata1995} explicitly discussed the undergoing merger between the Milky Way and the Sagittarius dwarf galaxy causing strong tidal disruptions due to its proximity to the Galactic Center and its radial velocity that would imply the transfer of its constituent matter, including gas molecules. Using the galactic fountain events first delineated by \cite{Shapiro1976, Bregman1980} in the Milky Way, star formation and supernova activity in a galaxy's disk drive cycles of gas ejection and re-accretion. This is also seen by \cite{Li2023} in NGC 2403 where the galactic fountains eject and contribute back to the disc-halo interface, which can then potentially drive the inside-out disk growth and redistribute its composition. Presence of galactic substructures such as the Radcliffe wave i.e. a massive undulating structure of gas and dust in the Milky Way \citep{Alves2020} which was seen to be oscillating and propagating radially outwards from the Galactic Center \citep{Konietzka2024} also supports the idea of migrating stars gathering prebiotic molecules before settling and prosper with life. 

\section{Conclusion} \label{sec:conclusion}

In this paper, we report the comprehensive observation of \htwoco{} absorption and \rrl{} emission in molecular clouds of the Galactic Plane sampled from the BPGS catalog. Using the Nanshan 25-m radio telescope, we identified 88 out of 215 sources exhibiting \htwoco{} absorption (40.93\%), with 59 of them being new detections. Only 11 sources (5.12\%) were detected with \rrl{} emission, and based on the \htwoco{} physical parameters, we confirm that the majority of the \bh{} objects are molecular clouds in the early stages of star formation. We also noted the double correlative nature of the \htwoco{} fluxes with infrared flux, which is thought to be a result of sub-CMB cooling. 

In context of the GHZ, cross-matched \bh{} objects show that the \htwoco{} absorption features are present across a wide range of galactocentric distances, primarily within the inner Galaxy (between $0.216$ and $10.769~\mathrm{kpc}$), which is an underexplored region in terms of presence of biogenic precursors. Detections of \htwoco{} in near proximity to the Galactic Center suggest sufficient shielding within dense molecular clouds to preserve COMs, and this challenges the traditional GHZ models that primarily rely on metallicity and catastrophic events as limiting factors. Our findings highlight the need to incorporate biogenic precursors (such as \htwoco{}) as a defining criterion of GHZ in the Milky Way. The negative-correlation observed between \htwoco{}'s detection fraction and column density with galactocentric distances also enhances the role of dynamic processes in redistributing prebiotically plausible molecules in the Galaxy.

While \htwoco{} alone may present some bias as a habitability tracer, its presence in conjunction with auxiliary key prebiotic molecules such as NH$_3$, HCN, and other NCHOPS molecules in general, may provide better outcomes as a GHZ indicator. Future studies should also incorporate physical limiters for habitability (e.g. AGN influence, supernova frequency, metallicity, stellar density) to further constraint the inner and outer radii of the GHZ. By expanding the chemical framework for defining the GHZ, our findings contribute to the broader effort of understanding the origins of life in the Universe and refining the search for prebiotic chemistry beyond our Earth.

\begin{acknowledgments}
We sincerely thank the staff of the Nanshan Observatory for their invaluable support during the observations. The observations for this study were conducted using the Nanshan 25-m Radio Telescope, which is operated by the Key Laboratory of Radio Astronomy, Chinese Academy of Sciences. The Nanshan 25-m Radio Telescope is partly supported by the Operation, Maintenence, and Upgrading Fund for Astronomical Telescopes and Facility Instruments, budgeted from the Ministry of Finance of China (MOF) and administered by the Chinese Academy of Sciences (CAS). We also extend our gratitude to the anonymous referee for their insightful comments and suggestions, which significantly improved the quality of this manuscript. 
\end{acknowledgments}

%

\vspace{5mm}
\facilities{XAO (Nanshan 25-m)}


\software{GILDAS \citep{Gildas2013}
          numpy \citep{Numpy2020}
          astropy \citep{TheAstropyCollaboration2022}
          matplotlib \citep{Matplotlib2007}
          }



\appendix

\section{Selected BGPS Source Parameters}

\begin{deluxetable}{cccccccc}
    \tablecaption{List of the BGPS sources observed using Nanshan 25-m.\label{tab:detection}}
    \colnumbers
    \tablehead{
        \colhead{Source} & \colhead{R.A. (J2000)} & \colhead{Decl. (J2000)} & \colhead{$l$} & \colhead{$b$} & \colhead{$S_{\mathrm{BGPS}}$} & \colhead{\htwoco{}} & \colhead{\rrl{}} \\
        & \colhead{(hh:mm:ss)} & \colhead{(dd:mm:ss.ss)} & \colhead{(deg)} & \colhead{(deg)} & \colhead{(Jy)} & \colhead{(Detected?)} & \colhead{(Detected?)}
    }
    \startdata
    BGPS0352 & 17:46:50 & -27:53:26.27 & 1.032 & 0.316 & 2.745 & N & N \\
    BGPS0512 & 17:48:56 & -27:44:37.00 & 1.398 & -0.006 & 2.282 & Y & N \\
    BGPS0523 & 17:47:48 & -27:33:25.14 & 1.428 & 0.306 & 2.963 & Y & N \\
    BGPS0548 & 17:49:57 & -27:44:14.28 & 1.518 & -0.194 & 2.941 & Y & N \\
    BGPS0598 & 17:50:55 & -27:40:25.07 & 1.684 & -0.348 & 1.170 & N & N \\
    \multicolumn{8}{c}{\vdots}
    \enddata
    \tablecomments{Columns: (1) BGPS source name; (2, 3) Coordinates in equatorial coordinate system; (4, 5) Coordinates in Galactic coordinate system; (6) BGPS $1.1~\mathrm{mm}$ continuum flux; (7, 8) \htwoco{} and \rrl{} detection flags i.e. 'Y' denotes detection and 'N' denotes non-detection. This table is available in its entirety in machine-readable form.}
\end{deluxetable}

\section{Observational Parameters of BGPS Sources with \htwoco{} Detection}

\begin{deluxetable}{ccccccc}
    \tablecaption{Observational parameters of the BGPS sources with \htwoco{} detection.\label{tab:observation}}
    \colnumbers
    \tablehead{
        \colhead{Source} & \colhead{Date} & \colhead{Elevation} & \colhead{$T_{\mathrm{sys}}$} & \colhead{$\sigma$} & \colhead{n(\htwoco{})} & \colhead{n(\rrl{})} \\
        & & \colhead{($^{\circ}$)} & \colhead{($\mathrm{K}$)} & \colhead{($\mathrm{K}$)} & &
    }
    \startdata
    BGPS0512 & 2013-08-14 16:05:15.000 & 15.3188 & 34.4313 & 0.0229 & 4 & 0 \\
    BGPS0523 & 2013-08-14 16:32:15.000 & 13.2965 & 35.6863 & 0.0234 & 2 & 0 \\
    BGPS0548 & 2013-08-14 16:59:15.000 & 10.8040 & 39.2719 & 0.0182 & 4 & 0 \\
    BGPS0647 & 2014-02-17 01:58:15.000 & 18.9745 & 22.1199 & 0.0128 & 3 & 0 \\
    BGPS0657 & 2014-02-17 02:24:55.000 & 19.1751 & 22.1199 & 0.0302 & 8 & 0 \\
    \multicolumn{7}{c}{\vdots}
    \enddata
    \tablecomments{Columns: (1) BGPS source name; (2) Observation date and time; (3) Observation elevation; (4) Antenna system temperature; (5) root-mean-square (RMS) noise; (6, 7) Number of \htwoco{} and \rrl{} lines detected. This table is available in its entirety in machine-readable form.}
\end{deluxetable}

\section{Gaussian Fitting Results of all \htwoco{} and \rrl{} Lines}

\begin{deluxetable}{ccCCCcc}
    \tablecaption{Gaussian fitting parameters of all \htwoco{} lines detected.\label{tab:h2co_params}}
    \colnumbers
    \tablehead{
        \colhead{Source} & \colhead{Brightness Temp.} & \colhead{Velocity} & \colhead{FWHM} & \colhead{Integrated Int.} & \colhead{Baseline RMS} & \colhead{Line RMS} \\
        & \colhead{(K)} & \colhead{($\mathrm{km}~\mathrm{s}^{-1}$)} & \colhead{($\mathrm{km}~\mathrm{s}^{-1}$)} & \colhead{($\mathrm{K}~\mathrm{km}~\mathrm{s}^{-1}$)} & \colhead{(K)} & \colhead{(K)}
    }
    \startdata
    BGPS0512 MC1 & -0.0578 & -53.9 \pm 0.4 & 4.0 \pm 1.0 & -0.25 \pm 0.05 & 0.0229 & 0.0197 \\
    BGPS0512 MC2 & -0.1930 & -27.53 \pm 0.31 & 17.1 \pm 0.7 & -3.50 \pm 0.24 & 0.0229 & 0.0197 \\
    BGPS0512 MC3 & -0.0833 & -2.4 \pm 1.6 & 38.4 \pm 2.7 & -3.41 \pm 0.27 & 0.0229 & 0.0197 \\
    BGPS0512 MC4 & -0.3350 & 79.41 \pm 0.22 & 42.4 \pm 0.6 & -15.15 \pm 0.16 & 0.0229 & 0.0197 \\
    BGPS0523 MC1 & -0.0365 & -2.8 \pm 1.6 & 26.1 \pm 3.2 & -1.01 \pm 0.12 & 0.0234 & 0.0181 \\
    \multicolumn{7}{c}{\vdots}
    \enddata
    \tablecomments{Columns: (1) BGPS source name with MC number; (2) Brightness temperature of the line; (3) Central velocity of the line; (4) FWHM of the line; (5) Integrated intensity of the line; (6, 7) Baseline and line RMS noise. This table is available in its entirety in machine-readable form.}
\end{deluxetable}

\begin{deluxetable}{ccCCCcc}
    \tablecaption{Gaussian fitting parameters of all \rrl{} lines detected.\label{tab:rrl_params}}
    \colnumbers
    \tablehead{
        \colhead{Source} & \colhead{Brightness Temp.} & \colhead{Velocity} & \colhead{FWHM} & \colhead{Integrated Int.} & \colhead{Baseline RMS} & \colhead{Line RMS} \\
        & \colhead{(K)} & \colhead{($\mathrm{km}~\mathrm{s}^{-1}$)} & \colhead{($\mathrm{km}~\mathrm{s}^{-1}$)} & \colhead{($\mathrm{K}~\mathrm{km}~\mathrm{s}^{-1}$)} & \colhead{(K)} & \colhead{(K)}
    }
    \startdata
    BGPS1326 HIIa & 0.0222 & 72.0 \pm 1.4 & 18.1 \pm 3.0 & 0.43 \pm 0.06 & 0.0150 & 0.0121 \\
    BGPS1520 HIIa & 0.0547 & 31.8 \pm 0.8 & 22.2 \pm 1.9 & 1.29 \pm 0.09 & 0.0192 & 0.0176 \\
    BGPS2045 HIIa & 0.0438 & 37.8 \pm 1.0 & 29.1 \pm 2.4 & 1.36 \pm 0.09 & 0.0162 & 0.0167 \\
    BGPS2094 HIIa & 0.0911 & 35.8 \pm 0.7 & 28.3 \pm 1.6 & 2.74 \pm 0.13 & 0.0232 & 0.0210 \\
    BGPS2313 HIIa & 0.0171 & 38.4 \pm 1.3 & 19.4 \pm 2.4 & 0.35 \pm 0.04 & 0.0101 & 0.0092 \\
    BGPS4244 HIIa & 0.0341 & 106.0 \pm 0.7 & 22.2 \pm 1.5 & 0.80 \pm 0.05 & 0.0100 & 0.0134 \\
    BGPS4372 HIIa & 0.0328 & 109.9 \pm 0.9 & 32.0 \pm 1.9 & 1.12 \pm 0.06 & 0.0101 & 0.0093 \\
    BGPS5385 HIIa & 0.0199 & 79.9 \pm 1.1 & 16.9 \pm 2.9 & 0.36 \pm 0.05 & 0.0111 & 0.0106 \\
    BGPS5884 HIIa & 0.0268 & 59.3 \pm 1.0 & 25.7 \pm 2.6 & 0.73 \pm 0.06 & 0.0101 & 0.0106 \\
    BGPS5910 HIIa & 0.0242 & 65.8 \pm 1.2 & 23.3 \pm 2.9 & 0.60 \pm 0.06 & 0.0104 & 0.0089 \\
    BGPS5910 HIIb & 0.0127 & 88.2 \pm 1.6 & 10.7 \pm 2.5 & 0.14 \pm 0.04 & 0.0104 & 0.0089 \\
    BGPS6661 HIIa & 0.0176 & -46.8 \pm 1.4 & 24.8 \pm 2.7 & 0.46 \pm 0.05 & 0.0099 & 0.0089 \\
    BGPS6661 HIIb & 0.0238 & 27.8 \pm 1.2 & 27.3 \pm 2.5 & 0.69 \pm 0.05 & 0.0099 & 0.0089 \\
    \enddata
    \tablecomments{Columns: (1) BGPS source name with \hii{} identifier; (2) Brightness temperature of the line; (3) Central velocity of the line; (4) FWHM of the line; (5) Integrated intensity of the line; (6, 7) Baseline and line RMS noise.}
\end{deluxetable}

\section{Master \bh{} Catalog}

\begin{deluxetable}{lLl}
    \tablecaption{Entries of the master \bh{} catalog.\label{tab:master}}
    \tablehead{
        \colhead{Entry} & \colhead{Units} & \colhead{Description}
    }
    \startdata
        BGPS & ... & BGPS source name \\
        RAdeg & \mathrm{deg} & Right ascension (J2000) \\
        DEdeg & \mathrm{deg} & Declination (J2000) \\
        $l$ & \mathrm{deg} & Galactic longitude \\
        $b$ & \mathrm{deg} & Galactic latitude \\
        $S_{\mathrm{BGPS}}$ & \mathrm{Jy} & BGPS $1.1~\mathrm{mm}$ continuum flux \\
        \htwoco{} & ... & \htwoco{} detection flag (`Y': detection, `N': non-detection) \\
        \rrl{} & ... & \rrl{} detection flag (`Y': detection, `N': non-detection) \\
        First Detection & ... & First detection flag (`Y': first detection, `N': previously detected) \\
        N(\htwoco{}) & ... & Number of \htwoco{} absorption lines detected \\
        N(\rrl{}) & ... & Number of \rrl{} emission lines detected \\
        $\mathrm{T}_{\mathrm{sys}}$ & \mathrm{K} & Antenna system temperature \\
        T(\htwoco{}) & \mathrm{K} & Brightness temperature of \htwoco{} line \\
        S(\htwoco{}) & \mathrm{Jy} & Flux density of \htwoco{} line \\
        V(\htwoco{}) & \mathrm{km}~\mathrm{s}^{-1} & Central velocity of \htwoco{} line \\
        FWHM(\htwoco{}) & \mathrm{km}~\mathrm{s}^{-1} & FWHM of \htwoco{} line \\
        $\mathrm{T}_{int}$(\htwoco{}) & \mathrm{K}~\mathrm{km}~\mathrm{s}^{-1} & Integrated intensity of \htwoco{} line \\
        T(\rrl{}) & \mathrm{K} & Brightness temperature of \rrl{} line \\
        S(\rrl{}) & \mathrm{Jy} & Flux density of \rrl{} line \\
        V(\rrl{}) & \mathrm{km}~\mathrm{s}^{-1} & Central velocity of \rrl{} line \\
        FWHM(\rrl{}) & \mathrm{km}~\mathrm{s}^{-1} & FWHM of \rrl{} line \\
        $\mathrm{T}_{\mathrm{int}}$(\rrl{}) & \mathrm{K}~\mathrm{km}~\mathrm{s}^{-1} & Integrated intensity of \rrl{} line \\
        $R_{\mathrm{gal}}$ & \mathrm{kpc} & Galactocentric distance based on cross-matching with \citetalias{MivilleDeschnes2017} \\
        $\mathrm{KD}$ & \mathrm{kpc} & Kinematic distance based on cross-matching with \citetalias{MivilleDeschnes2017} \\
        $z_{\mathrm{gal}}$ & \mathrm{kpc} & Distance to Galactic midplane based on cross-matching with \citetalias{MivilleDeschnes2017} \\
        V($^{12}$CO) & \mathrm{km}~\mathrm{s}^{-1} & Central velocity of $^{12}$CO line based on cross-matching with \citetalias{MivilleDeschnes2017} \\
        $\tau_{\mathrm{app}}$ & ... & Derived apparent optical depth based on \htwoco{} line \\
        $N(\rm\htwoco{})$ & \mathrm{cm}^{-2} & Derived \htwoco{} column density based on \htwoco{} line
    \enddata
    \tablecomments{The full master table includes all the selected BGPS sources and other information including \htwoco{} parameters, \rrl{} parameters, \citetalias{MivilleDeschnes2017} cross-matched parameters, and derived \htwoco{} parameters. If the data is unavailable, the columns will be masked appropriately. Spectral line results and derived parameters are accompanied with their standard deviations. This table is available in its entirety in machine-readable form.}
\end{deluxetable}


\bibliography{ref}{}
\bibliographystyle{aasjournal}



\end{document}